\documentclass[aps,prb,twocolumn,groupedaddress,showpacs]{revtex4}
\usepackage{amsmath,amssymb,amsfonts}
\usepackage{graphicx}

\begin{document}

\title{Topographical fingerprints of many-body interference\\
in STM junctions on thin insulating films}

\author{Andrea Donarini}
\email[]{andrea.donarini@physik.uni-r.de}
\author{Sandra Sobczyk}
\author{Benjamin Siegert}
\author{Milena Grifoni}

\affiliation{Institut f\"{u}r Theoretische Physik, Universit\"at Regensburg, D-93040 Regensburg, Germany}

\date{\today}

\begin{abstract}
Negative differential conductance (NDC) is a non-linear transport phenomenon ubiquitous in molecular nanojunctions. Its physical origin can be the most diverse. In rotationally symmetric molecules with orbitally degenerate many-body states it can be ascribed to interference effects. We establish in this paper a criterion to identify the interference blocking scenario by correlating the spectral and the topographical information achievable in an STM single molecule measurement. Simulations of current voltage characteristics as well as constant height and constant current STM images for a Cu-Phthalocyanine (CuPc) on a thin insulating film are presented as experimentally relevant examples.
\end{abstract}

\maketitle

\section{INTRODUCTION}

% \noindent \emph{Introduction} --
Negative differential conductance (NDC) is a fundamental property of two terminal devices since the discovery of the first tunnel diode \cite{Esaki58}. The realization of NDC within an atomic scale device \cite{LyoA89,BedrossianCMG_89,ChenRRT_99,ZengWWYH_00,RinkioJKT_10, FrankeSHFPZRCL_08} can consequently be regarded as a milestone in the process of miniaturization which drives the information technology.

Scanning tunneling microscopy (STM) experiments have played an important role in this research field giving several examples of NDC observed with a variety of nanojunctions. A number of physical scenarios have been proposed for the explanation of the experimental findings: among others the existence of sharp resonances on both electrodes \cite{BedrossianCMG_89, XueDHRHK_99}, the voltage dependent increase in the tunneling barrier height \cite{GrobisWYC_05, TuMH_08}, the orbital matching between molecule and tip \cite{ChenHZWLYH_07, ShiPXCZMvH_09} or even just the symmetry matching between surface states in the substrate and molecular states \cite{HeinrichRCFL_11}. Last but not least vibrational mediated NDC has also been observed in single molecule devices \cite{GaudiosoLH_00} and proposed to test position dependent Franck-Condon factors in suspended carbon nanotubes \cite{TraversoPCS_11}.

Recently also interference phenomena in single molecule junctions have attracted intense theoretical \cite{CardamoneSM_06, KeB_08, QianLZHS_08, BegemannDDG_08, SolomonAHGWDR_08, DarauBDG_09, DonariniBG_09, DonariniBG_10, MarkussenST_10, TsujiSY_11, MarkussenST_11, Ernzerhof_11} and experimental \cite{MayorWREvHBF_03, TaniguchiTMSTYK_11, AradhyaMKAPSNV_12, GuedonVMTHM_12} investigations. These junctions allow to tackle the fundamental question of the quantum mechanical nature of the electronic transport at the nanoscale and exhibit dramatic modulations of the current desirable for applications. The quest of specific fingerprints of the electronic interference which go beyond the bare current or conductance suppression \cite{AradhyaMKAPSNV_12} remains, though, a crucial issue. We establish in this article a criterion to identify the interference blocking scenario by correlating the spectral and the topographical information achievable in an STM single molecule measurement.

In a recent publication we have predicted the occurrence of NDC due to interference blocking \cite{DarauBDG_09,DonariniBG_09,DonariniBG_10} in an STM single benzene junction on thin insulating film\cite{SobczykDG_12}. Benzene, however is not easily accessible in STM experiments and it is not obvious to which extent the findings of [\onlinecite{SobczykDG_12}] apply to larger, experimentally relevant molecules, see {\it e.g.} [\onlinecite{ReppMSGJ_05,LiljerothRM_07}]. A major result of this article is an analytical expression for the current as a function of the applied bias voltage encompassing various transport regimes, see Eq. \eqref{eq:Current} and \eqref{eq:Populations}, which provides both the criteria for the occurrence of interference blocking NDC and the interpretation of its topographical fingerprints.
%
%%%%%%%%%%%%%%%%%%%%%%%%%%%%%%%%%%%%%%%%%%%%%%%%%%%%%%%%%%%%%%%%
% Figure 1
%%%%%%%%%%%%%%%%%%%%%%%%%%%%%%%%%%%%%%%%%%%%%%%%%%%%%%%%%%%%%%%%
 \begin{figure}[h!]
 \includegraphics[width = 0.75\columnwidth]{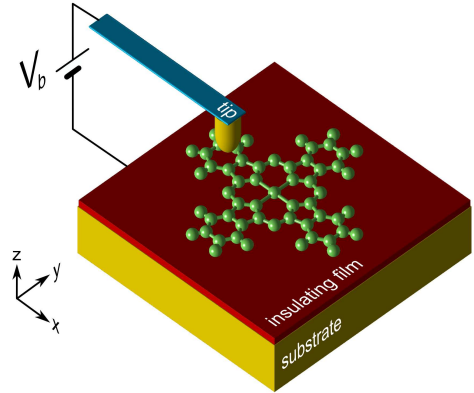}
 \caption{(Color online) Artistic view of an STM single molecule junction. We show in yellow (light gray) the metallic leads (tip and substrate), in red (dark gray) the thin insulating film and in green (middle gray) the schematic representation of a CuPc.}\label{fig:Set-up}
 \end{figure}
%%%%%%%%%%%%%%%%%%%%%%%%%%%%%%%%%%%%%%%%%%%%%%%%%%%%%%%%%%%%%%%
%
Specifically, interference NDC is expected to occur in molecules which exhibit an electron affinity (ionization potential) $E_0-E_1$ ($E_{-1} - E_0$) very close to the work function $\phi_0$ of the substrate and, due to their rotational symmetry, have an orbitally degenerate anion (cation) many-body eigenstate (here $E_0$, $E_{\pm 1}$ denote the many-body ground state energy of the neutral molecule and of the anion or cation). The necessary decoupling from the substrate, originally obtained through a thin insulating layer \cite{ReppMSGJ_05}, can also be achieved by combining two different molecules in a monolayer directly adsorbed on a metal surface \cite{FrankeSHFPZRCL_08}. Recently, this set-up has been used to demonstrate position dependent local gating \cite{TorrenteKSFP_12}, thus suggesting an alternative possibility to achieve the mentioned interference conditions. Topographical fingerprints to identify the interference blocking scenario are predicted both for the constant height and constant current scanning modes. Interference is associated in the first case to a flattening of the current map in the 
molecule region with a corresponding loss of the characteristic nodal plane pattern (see Fig.~\ref{fig:Current_maps}); in the second case to an enhanced sensitivity of the apparent height of the molecule to the operating current (see Fig.~\ref{fig:Height_cuts}). As we will prove later, both phenomena have the same origin {\it i.e.}, in the interference blocking regime, the bottleneck process defining the current pattern is a \emph{substrate} and not a tip tunneling event. The analytical results apply to a wide class of molecular junctions. In particular, we present simulations concerning a CuPc junction.

\section{Model}

% \noindent \emph{Model} --
We describe the STM single molecule junction as a system-bath model:
\begin{equation}
H = H_{\mathrm{mol}} + H_{\mathrm{sub}} + H_{\mathrm{tip}} + H_{\mathrm{tun}},
\end{equation}
where $H_{\mathrm{mol}}$ is the Hamiltonian for the isolated molecule, in which, to fix the ideas, we distinguish a single particle component $H_0$ and a two particle component $V$, both expressed in terms of creation and annihilation operators $\{ d_{\alpha \sigma}, d^\dagger_{\alpha\sigma} \}$ for the atomic orbitals $\psi_\alpha$ where $\alpha$ indicates both the site and the atomic species.  $H_{\mathrm{sub}}$ and $H_{\mathrm{tip}}$ account for the substrate and the tip, respectively, which we assume as reservoirs of non interacting electrons with different spatial confinement. Finally $H_{\mathrm{tun}}$ describes the tunneling coupling between the metallic leads and the molecule:
\begin{equation}
\label{eq:H_tun}
H_{\mathrm{tun}} = \sum_{\chi k\ell m\sigma}
t^{\chi}_{{k}\ell m}c^\dag_{{\chi k}\sigma}d_{\ell m\sigma}
+ h.c.
\end{equation}
where $\chi = S,T$ indicates the substrate or the tip, $k$ the momentum and $\sigma$ the spin of the electron in the lead. Due to their rotational symmetry, the molecular orbitals are classified using the projection $\ell$ of the angular momentum along the principal rotation axis of the molecule. A further quantum number $m$ is introduced to account for possible degeneracies in the spectrum of the angular momentum. Finally, the tunneling amplitudes $t^{\chi}_{{k}\ell m}$ take the form
\begin{equation}
\begin{split}
t^{\mathrm S}_{\vec{ k} \ell m} &=
\varepsilon_{\ell m }
\langle S \vec{k} \sigma | \ell m \sigma \rangle,\\
t^{\mathrm T}_{k_z \ell m} &=
\varepsilon_{\ell m}
\langle T k_z \sigma | \ell m \sigma \rangle,
\end{split}
\end{equation}
where $\varepsilon_{\ell m}$ is the energy eigenvalue of the single particle Hamiltonian $H_0$ associated to the state  $|\ell m \sigma\rangle$. For the substrate and tip states we assume the model described in [\onlinecite{SobczykDG_12}]: a three dimensional momentum is necessary for the extended substrate states while only the momentum in the transport direction characterizes the tip states which are confined in the $x$ and $y$ direction around the tip position.

%%%%%%%%%%%%%%%%%%%%%%%%%%%%%%%%%%%%%%%%%%%%%%%%%%%%%%%%%%%%%%%%
% Figure 2
%%%%%%%%%%%%%%%%%%%%%%%%%%%%%%%%%%%%%%%%%%%%%%%%%%%%%%%%%%%%%%%%
 \begin{figure}[t]
 \includegraphics[width = \columnwidth]{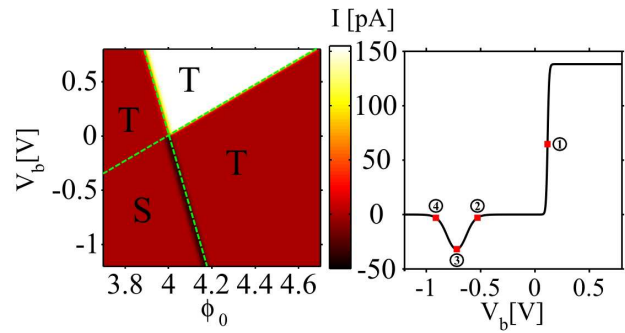}
 \caption{(Color online) Left panel: Current through a CuPc single molecule junction as a function of the substrate (and tip) work function $\phi_0$ and of the sample bias $V_b$. The tip apex position is assumed at $(x,y,z-d) = (+5,-5,7)$\AA~ with the origin taken on the metal-insulator interface and in correspondance of the center of the molecule, and $d$ being the thikness of the insulating layer (see Fig.~\ref{fig:Set-up}). The tip and substrate resonant lines (respectively with positive and negative slopes) divide the parameter space into four regions. T (S) indicates a region in which the current is proportional to the tip (substrate) tunneling rate. Right panel: Current obtained from a cut of the left panel plot corresponding to $\phi_0 = 4.1 eV$. The numbers on the current-voltage plot refer to the current maps of Fig.~\ref{fig:Current_maps}. The current scale is the same for the left and right panel.}
 \label{fig:IV_stability}.
 \end{figure}
%%%%%%%%%%%%%%%%%%%%%%%%%%%%%%%%%%%%%%%%%%%%%%%%%%%%%%%%%%%%%%%

Our method of choice to treat the dynamics in the regime of weak coupling between system and leads is the Liouville equation method.
We start from the Liouville equation for the total density operator $\rho(t)$ of the whole system consisting of the molecule, the tip and the substrate. We focus on the time evolution of the reduced density matrix  $\sigma={\mathrm Tr}_{\mathrm S+T}\{\rho\}$, formally obtained by taking the trace over the unobserved degrees of freedom of the  tip and the substrate. A detailed discussion and derivation of the equation of motion for the reduced density operator of the system can be found {\it e.g.} in \cite{Blum_book, DarauBDG_09} and in \cite{SobczykDG_12} its adaptation to the STM set-up on thin insulating films. For a general discussion about the reduced density matrix and related equations of motion see also \cite{Weiss_12, GrifoniSW_96}

%%%%%%%%%%%%%%%%%%%%%%%%%%%%%%%%%%%%%%%%%%%%%%%%%%%%%%%%%%%%%%%%
% Figure 3
%%%%%%%%%%%%%%%%%%%%%%%%%%%%%%%%%%%%%%%%%%%%%%%%%%%%%%%%%%%%%%%%
 \begin{figure}[h!]
 \includegraphics[width =\columnwidth]{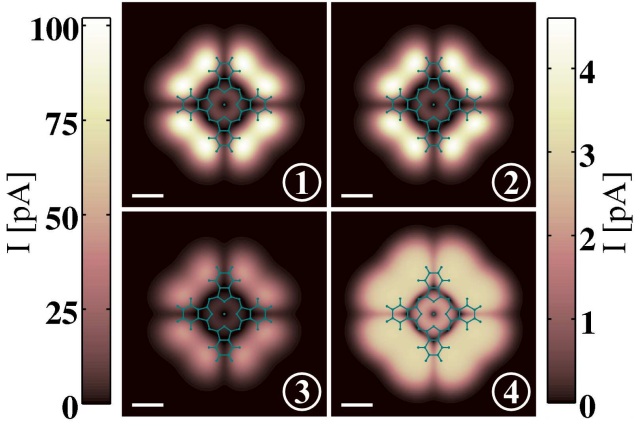}
 \caption{Constant height current maps calculated for different bias voltages. The color bar on the left (right) hand side corresponds to the maps 1 and 3 (2 and 4). The 5\AA~long white line sets the scale of the images. The numbers in the maps refer to the biases indicated in the right panel of Fig. \ref{fig:IV_stability}. The current map in the interference blockade regime (map 4) appears flat in the molecule region. The characteristic nodal planes pattern appears instead much more pronounced at the positive and negative bias resonances (map 1 and 3) and even in the Coulomb blockade region (map 2). The tip apex is placed at 7\AA~above the molecular plane while the substrate biases are, respectively $V_{b1} = 0.1153$V, $V_{b2} = -0.5303$V, $V_{b3} = -0.7201$V, and $V_{b4} = -0.9118$V.}
 \label{fig:Current_maps}
 \end{figure}
%%%%%%%%%%%%%%%%%%%%%%%%%%%%%%%%%%%%%%%%%%%%%%%%%%%%%%%%%%%%%%%

\begin{figure}[h]
 \includegraphics[width = \columnwidth]{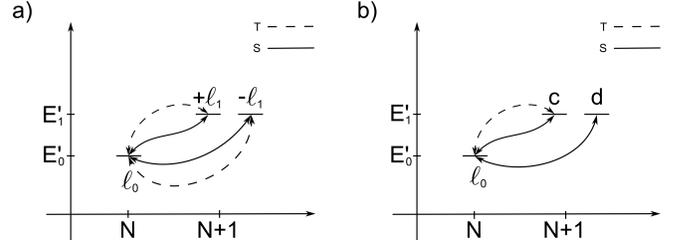}
 \caption{Schematic representation of the many-body states participating to the transport. On the vertical axis we report the grand canonical energies $E'_0 := E_0 - {\rm N}\mu_0$ and $E'_1 := E_1 - ({\rm N}+1)\mu_0$, being $\mu_0$ the equilibrium chemical potential for the leads. In panel a) we adopt the angular momentum representation while in panel b) the decoupling basis is introduced for the anionic states (see main text for details).}
 \label{fig:Rate_Schemes}
 \end{figure}

Let us now consider a molecule deposited on a substrate having a non degenerate neutral ground state $|{\rm N}\,E_0\,\ell = \ell_0\,S_z = 0\rangle$ and an orbitally degenerate anion ground state $|{\rm N}+1\,E_1\,\ell = \pm\ell_1\,S_z = \pm1/2\rangle$. Assume that the neutral state minimizes the grand canonical Hamiltonian $H_{\mathrm G} = H-\mu_0N$, where $\mu_0 = -\phi_0$ is the equilibrium chemical potential of the leads, and $\phi_0$ the corresponding work function. If $E_1 - E_0 \approx \mu_0$ there is a bias window in which the transport characteristics are dominated by a dynamics which involves the neutral and anionic ground states only. In panel a) of Fig.~\ref{fig:Rate_Schemes} we give a schematic representation of the many body states participating in the transport and the associated transition rates where, for the sake of simplicity, we neglect the spin degree of freedom. According to the general theory presented in [\onlinecite{SobczykDG_12}], the corresponding generalized master equation for the reduced density matrix in the angular momentum basis reads:

\begin{equation}
\label{eq:GME}
\begin{split}
\dot{\sigma}^{{\rm N}E_0}_{\ell_0\ell_0} &=
-\sum_{\chi \tau \ell}
R^{\chi \tau}_{\ell-\ell_0,\, \ell-\ell_0}(\Delta E)f^+_\chi(\Delta E) \sigma^{{\rm N}E_0}_{\ell_0\ell_0}\\
&+ \sum_{\chi \tau \ell \ell'}
R^{\chi \tau}_{\ell-\ell_0,\, \ell'-\ell_0}(\Delta E)
f^-_\chi(\Delta E)
\sigma^{{\rm N}+1E_1 \tau}_{\ell'\ell}\\
\dot{\sigma}^{{\rm N}+1E_1\tau}_{\ell\ell'} &=
-\frac{1}{2}\sum_{\chi \ell''}
\Big[
R^{\chi \tau}_{\ell-\ell_0,\, \ell''-\ell_0}(\Delta E)
\sigma^{{\rm N}+1E_1 \tau}_{\ell'' \ell'},\\
&
\sigma^{{\rm N}+1E_1 \tau}_{\ell \ell''}
R^{\chi \tau}_{\ell''-\ell_0,\, \ell'-\ell_0}(\Delta E)
\Big]f^-_{\chi}(\Delta E)\\
&+ \sum_{\chi \tau}
R^{\chi \tau}_{\ell-\ell_0,\, \ell'-\ell_0}(\Delta E)
f^+_\chi(\Delta E)
\sigma^{{\rm N} E_0}_{\ell_0\ell_0},\\
\end{split}
\end{equation}
where $\ell,\ell'$ and $\ell'' = \pm \ell_1$, span the angular momenta of the anionic ground state and $\Delta E = E_1-E_0$ is the energy difference between the anionic and neutral ground states. Moreover $f^+_\chi(x)$ is the Fermi function for the lead $\chi$, $f^+_\chi(x):= f(x-\mu_\chi)$ and $f^-_{\chi}(x):= 1 - f^+_\chi(x)$. Note that we assume an asymmetric potential drop where $\mu_{T} = \mu_0 - ceV_b$ with $c = 0.87$ and $\mu_S - \mu_T = eV_b$. The rate $R^{\chi \tau}_{\ell-\ell_0,\, \ell'-\ell_0}$ is defined as:

\begin{equation}
\begin{split}
R^{\chi \tau}_{\Delta\ell,\,\Delta\ell'} (\Delta E)=&
\sum_{mm'}
\langle {\rm N}+1 E_1 \ell \tau |
d^\dagger_{\Delta \ell\, m \tau}
| {\rm N} E_0 \ell_0 0 \rangle\\
&\times
\Gamma^\chi_{\Delta\ell m,\Delta\ell' m'}(\Delta E)\\
&\times
\langle {\rm N} E_0 \ell_0 0|
d_{\Delta\ell'\, m' \tau}
|{\rm N}+1 E_1 \ell' \tau \rangle,
\end{split}
\end{equation}
where
\begin{equation}
\Gamma_{\Delta \ell m,\,\Delta\ell' m'}^{\chi}(\Delta E) = \frac{2\pi}{\hbar}
\sum_k
\left(t^{\chi}_{k \Delta \ell m}\right)^*
t^{\chi}_{k \Delta \ell' m'}
\delta(\varepsilon_k^{\chi}-\Delta E),
\end{equation}
and we have introduced the notation $\Delta\ell = \ell - \ell_0$, $\Delta\ell' = \ell' - \ell_0$ for the variation in angular momenta associated to the tunneling process.

Due to the rotational symmetry of the molecule and the different spatial confinement of the leads, the rate matrices acquire the form:
\begin{equation}
\label{eq:Rate_Mat}
\begin{split}
R^{S}_{\Delta\ell,\, \Delta\ell'} & = R^S
\delta_{\Delta\ell,\, \Delta\ell'},\\
R^{T}_{\Delta\ell,\, \Delta\ell'}& = R^T
\exp\left(
{-\mathrm i}
\frac{\Delta\ell-\Delta\ell'}{\Delta\ell}\,
\phi_{\Delta\ell}
\right),
\end{split}
\end{equation}
where we did not write for simplicity the energy dependence of $R^S$ and the energy and tip position dependence of $R^T$ and of the phase $\phi_{\Delta\ell}$. Moreover, the latter is defined as
\begin{equation}
\phi_{\Delta\ell} =
{\mathrm{arg}}
\left(
\sum_m t^T_{\tilde{k}\Delta\ell m} \langle {\rm N} E_0 \ell_0 0|
d_{\Delta\ell m\tau}|{\rm N}+1 E_1 \ell \tau \rangle
\right).
\end{equation}

Due to their particular structure, the rate matrices \eqref{eq:Rate_Mat} are both diagonalized by the same basis transformation. 
%%%%%%%%%%%%%%%%%%%%%%%%%%%%%%%%%%%%%%%%%%%%%%%%%%%%%%%%%%%%%%%%
% Figure 4
%%%%%%%%%%%%%%%%%%%%%%%%%%%%%%%%%%%%%%%%%%%%%%%%%%%%%%%%%%%%%%%%
%%%%%%%%%%%%%%%%%%%%%%%%%%%%%%%%%%%%%%%%%%%%%%%%%%%%%%%%%%%%%%%

While the substrate rate matrix is invariant under whatever unitary transformation, the tip rate matrix acquires a peculiar diagonal form since one of its eigenvalues vanishes. The basis transformation, within each spin sector of the anionic ground state, reads
\begin{equation}
\left(
  \begin{array}{c}
    |{\mathrm c}\rangle \\
    |{\mathrm d} \rangle \\
  \end{array}
\right)
= \frac{1}{\sqrt{2}}
\left(
  \begin{array}{cc}
    e^{-{\mathrm i}\phi_{\Delta\ell}} & e^{+{\mathrm i}\phi_{\Delta\ell}} \\
    e^{-{\mathrm i}\phi_{\Delta\ell}} & -e^{+{\mathrm i}\phi_{\Delta\ell}} \\
  \end{array}
\right)
\left(
  \begin{array}{c}
    |+\!\ell_1 \rangle \\
    |-\!\ell_1 \rangle \\
  \end{array}
\right)
\end{equation}
and it depends on the position of the tip via the phase $\phi_{\Delta\ell}$. Due to the diagonal form of the rate matrices, in this basis the dynamics is described only by means of populations. In particular the \emph{decoupled} states $|{\rm N}+1 \, E_1\, {\mathrm d} \, \tau \rangle$ are only coupled to the neutral ground state $|{\rm N} \, E_0 \, \ell_0 \, 0 \rangle $ via substrate-molecule tunneling events. Both tunneling couplings are still open instead for the \emph{coupled} states $|{\rm N}+1 \, E_1 \, {\mathrm c} \, \tau \rangle$, see panel b) in Fig.~\ref{fig:Rate_Schemes}. The corresponding master equation reads:

\begin{widetext}
\begin{equation}\label{eq:ME}
\left(
\begin{array}{c}
  \dot{\sigma}^{{\rm N}} \\
  \dot{\sigma}^{{\rm N}+1 \tau}_{\mathrm c} \\
  \dot{\sigma}^{{\rm N}+1 \tau}_{\mathrm d}
\end{array}
\right)
=
\left[
2R^T
\left(
  \begin{array}{ccc}
    -2f_T^+ & 2f_T^- & 0 \\
    f_T^+ & -f_T^- & 0 \\
    0 & 0 & 0 \\
  \end{array}
\right)
+
R^S
\left(
  \begin{array}{ccc}
    -4f_S^+ & 2f_S^- & 2f_S^- \\
    f_S^+ & -f_S^- & 0 \\
    f_S^+ & 0 & -f_S^- \\
  \end{array}
\right)
\right]
\left(
\begin{array}{c}
  \sigma^{{\rm N}} \\
  \sigma^{{\rm N}+1 \tau}_{\mathrm c} \\
  \sigma^{{\rm N}+1 \tau}_{\mathrm d}
\end{array}
\right),
\end{equation}
\end{widetext}
where for simplicity we have omitted the arguments ($\Delta E$) of the Fermi functions and the tunneling rates $R^\chi$ and suppressed the indexes $E_0$, $\ell_0$ and $E_1$ in the elements of the density matrix. The stationary current flowing through the STM junction is  calculated as the average $\langle I_S \rangle = {\mathrm{Tr}}\{\sigma^{\mathrm{stat}} I_S\} = -\langle I_T \rangle$ where $\sigma^{\mathrm{stat}}$ is the stationary solution of Eq.~\eqref{eq:ME} and $I_\chi$  are the current operators which are directly obtained from Eq.~\eqref{eq:GME} following, for example, \cite{DarauBDG_09, SobczykDG_12}. Despite its simplicity, Eq.~\eqref{eq:ME} describes the system in a variety of different regimes which leave their fingerprints in the current voltage characteristics and current maps.

\section{Results}

% \noindent \emph{Results} --
Given Eq.~\eqref{eq:ME}, the stationary current flowing through the system is found in closed analytical form, what represents one major result of this work. It reads:

\begin{equation}\label{eq:Current}
I(\vec{R}_{\mathrm{tip}},V_{\mathrm{b}}) = 2eR^Sf^+_S \sigma^{\rm N}
\left(
1-\frac{\sigma^{{\rm N}+1 \tau}_{\mathrm c}}{\sigma^{{\rm N}+1 \tau}_{\mathrm d}}
\right)
\end{equation}
where $e$ is the (negative) electron charge and
\begin{equation}
\label{eq:Populations}
\begin{split}
\sigma^{\rm N} &=
\left(
1+
2 \frac{R^S f^+_S + 2R^T f^+_T}
{R^S f^-_S + 2R^T f^-_T} +
2 \frac{f^+_S}{f^-_S}
\right)^{-1},\\
\frac{\sigma^{{\rm N}+1 \tau}_{\mathrm c}}{\sigma^{{\rm N}+1 \tau}_{\mathrm d}} &=
\frac{R^S f^+_S + 2R^T f^+_T}
{R^S f^-_S + 2R^T f^-_T}
\cdot\frac{f^-_S}{f^+_S}.
\end{split}
\end{equation}

Depending the rate $R^T$  on the tip position and the bias and the Fermi functions on the bias, both topographical and spectral information is embedded in Eq.~\eqref{eq:Current}.

In the right panel of Fig.~\ref{fig:IV_stability} we report the IV characteristics calculated for a Cu-Phthalocyanine on a metal-insulator substrate (a 7\AA~thick insulator with relative dielectric constant $\varepsilon_{\mathrm r} = 5.9$) with an effective work function $\phi_S = 4.1 eV$. We set up the single particle Hamiltonian for the molecule in the tight binding approximation and calculate the hopping terms following the Slater-Koster scheme \cite{SlaterK_54}. Moreover, we adopt the constant interaction approximation  and assume a charging energy that fits the experimentally evaluated electron affinity $E_0-E_1$ of CuPc of $4 eV$.

At low bias the current is suppressed by Coulomb blockade. As the bias increases on the positive side (conventionally under this condition electrons flow  from the tip to the substrate) the current undergoes a sudden jump corresponding to the opening of the neutral-anion transition at the tip-molecule interface ($E_1-E_0 = \mu_T$). On the negative bias side the Coulomb blockade is also lifted, but this time at the substrate resonance point ($E_1-E_0 = \mu_S$) and the current shows a sharp peak whose width scales with the temperature ($k_{\mathrm B}T = 6 meV$ in all presented plots). At higher negative biases the current is blocked due to interference and the decoupled anionic state is the sink of the system. A crucial condition for the interference blocking to occur is that $E_1-E_0 \ll E_0-E_{-1}$, ensuring that the substrate-molecule anion resonance anticipates the tip-molecule cation one which would otherwise dominate the transport characteristics.

Analogous interference blocking involving degenerate manybody states has been encountered in a variety of systems \cite{DarauBDG_09,DonariniBG_09,DonariniBG_10,BegemannDDG_08,SobczykDG_12}. Nevertheless the STM set up described here uniquely allows to correlate the interference current blocking with specific topographical fingerprints. In Fig.~\ref{fig:Current_maps} we present different constant height current maps (the tip is positioned always 7\AA~above the molecular plane) corresponding to the different points labeled in the right panel of Fig. \ref{fig:IV_stability}. Maps 1 and 3 are calculated for the tip and substrate resonant tunneling conditions while maps 2 and 4 for the Coulomb and interference blockade regimes, respectively. Striking is the flattening of the current map obtained in the interference case (map 4) if compared to all other regimes. 
%%%%%%%%%%%%%%%%%%%%%%%%%%%%%%%%%%%%%%%%%%%%%%%%%%%%%%%%%%%%%%%%
% Figure 5
%%%%%%%%%%%%%%%%%%%%%%%%%%%%%%%%%%%%%%%%%%%%%%%%%%%%%%%%%%%%%%%%
 \begin{figure}[h]
 \includegraphics[width = 0.9\columnwidth]{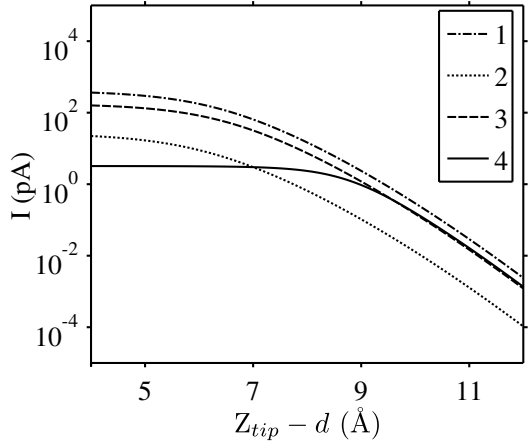}
 \caption{Current vs tip-molecule distance calculated for different biases. The numbers in the legend correspond to the different cases illustrated in Fig.~\ref{fig:Current_maps}: respectively $V_{b1} = 0.1153$V, $V_{b2} = -0.5303$V, $V_{b3} = -0.7201$V and $V_{b4} = -0.9118$V.  Notice in particular the wide plateau associated to the interference blockade regime (line 4) and its crossing with the Coulomb blockade line for $Z_{tip}-d = 7$\AA.}
 \label{fig:Current_z}
 \end{figure}
%%%%%%%%%%%%%%%%%%%%%%%%%%%%%%%%%%%%%%%%%%%%%%%%%%%%%%%%%%%%%%%
Signatures of interference can be clearly seen also in the current vs. tip-molecule distance represented in Fig.~\ref{fig:Current_z}. The four traces correspond to the four different biases conditions indicated with the numbers $1$ to $4$ in the right panel of Fig.~\ref{fig:IV_stability} and the tip is in the same $xy$ position. At large tip-molecule distances all traces show the exponentially decaying behaviour typical of the STM measurements (roughly 1 order of magnitude decay per \AA). At shorter distances, all curves saturates due to the form of the $p_z$ orbitals. Contrary to the others, though, the curve corresponding to the interference blockade regime (case 4) saturates at larger distances and shows a wide plateau. For this reason it even crosses the Coulomb blockade trace (case 2) at $\Delta z = 7$\AA, consistently with the result of Fig. \ref{fig:IV_stability}. 
%%%%%%%%%%%%%%%%%%%%%%%%%%%%%%%%%%%%%%%%%%%%%%%%%%%%%%%%%%%%%%%%
% Figure 6
%%%%%%%%%%%%%%%%%%%%%%%%%%%%%%%%%%%%%%%%%%%%%%%%%%%%%%%%%%%%%%%%
 \begin{figure}[h]
 \includegraphics[width = 0.9\columnwidth]{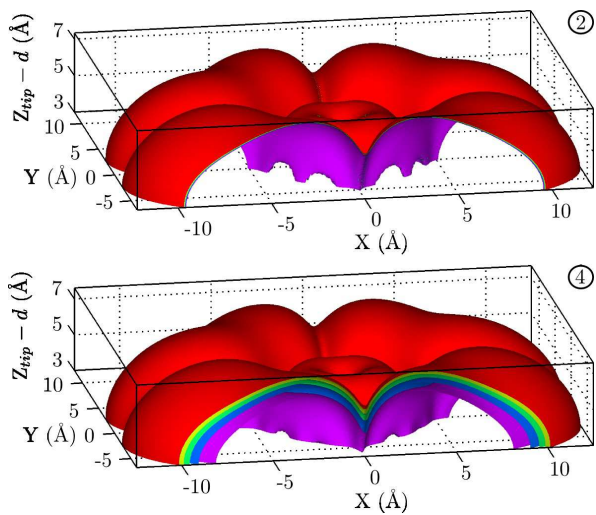}
 \caption{(Color online) Isosurfaces of constant current calculated in the proximity of the Coulomb blockade (upper panel, $V_b = -0.5303\,$V) and interference blockade (lower panel, $V_b = -0.9118\,$V) regimes. The surfaces correspond in both cases to the currents: $I = 3.15$, $3.075$, $3.0$, $2.925$, $2.85\,$pA.}
 \label{fig:Height_cuts}
 \end{figure}
%%%%%%%%%%%%%%%%%%%%%%%%%%%%%%%%%%%%%%%%%%%%%%%%%%%%%%%%%%%%%%%
Finally, we also present in Fig.~\ref{fig:Height_cuts} several constant current topographic maps simulated for different biases and different working currents. The surfaces presented in the upper panel correspond to the Coulomb blockade regime, while the ones in the lower panel to the interference blockade. Due to the particular choice of the biases, the apparent height of the molecule is exactly the same if we choose 3 pA as a working current. The shape of the molecule is not modified in the interference blockade regime, as it is for the constant height current maps (see Fig.\ref{fig:Current_maps}). Yet, striking it is, in this regime, the enhanced sensitivity of the apparent height of the molecule with respect to the variation of the working current, if compared with the same measurement  in the Coulomb blockade regime. The surfaces presented in Fig.~\ref{fig:Height_cuts} correspond in fact, for both cases, to working currents in the range 2.85 - 3.15 pA.

\section{Discussion}

All the results presented in the previous section can be understood by analyzing the different limits of Eq.~\eqref{eq:Current}. Let us first consider the Coulomb blockade regime. The latter is defined, for $V_{\mathrm b} < 0 $, by the inequality $E_1 - E_0 -\mu_{\mathrm S} \gg k_{\mathrm B}T$ which in turn implies $f_{\mathrm S}^+ \ll 1$ and $f_{\mathrm T}^+ \ll 1$. In this limit and under the asymmetry relation $R^T \ll R^S$ typical of an STM experiment, it is not difficult to prove that 

\begin{equation}
I_{\rm CB} = 4eR^Tf_T^-\frac{f_S^+}{f_S^-}\left( 1 + 4\frac{f_S^+}{f_S^-} \right)^{-1} \approx 4eR^T f_S^+.
\label{eq:I_CB}
\end{equation}

The current is thus proportional to the tip rate. The equality in Eq.~\eqref{eq:I_CB} has also a precise physical interpretation. The charge fluctuations at the substrate lead represent the fastest phenomenon ($f_S^+/f_T^+ \gg 1$ due to the asymmetric potential drop at tip-molecule and substrate-molecule contacts) which sets the ratio between the populations of the states to be the thermal average, $\sigma^{{\rm N}+1\tau}_{\rm c/d}/\sigma^N = f^+_S/f^-_S$. Finally, the trace sum rule implies:

\begin{equation}
 \sigma^N = \left( 1 + 4\frac{f_S^+}{f_S^-} \right)^{-1}.
\label{eq:Pop_CB}
\end{equation}

The current is determined instead by the  slowest process: the tunnelling event $|{\rm N}+1 \, E_1 \, {\mathrm c} \, \tau \rangle \to |{\rm N} \, E_0 \, 0\rangle$ towards  the tip. Equation    \eqref{eq:I_CB} follows due to the presence of $2$ spin channels and that the tip rate for the coupled state is $2 R^T$.  Analogously, for $V_{\mathrm b}>0$, the Coulomb blockade condition reads $E_1-E_0 - \mu_T \gg k_{\mathrm B}T$ and the current is again proportional to the tip rate, namely $I = -4eR^Tf_T^+$. Thus, the constant height current map reproduces the shape of the moelcular orbital encoded in $R^T$.

The interference blockade regime is confined to the negative bias and it is defined by the inequality $E_1 - E_0 -\mu_S \ll -k_{\mathrm B}T$ which implies $f^+_S \approx 1$ and $f^+_T \ll 1$. Under these conditions the current, Eq.~ \eqref{eq:Current}, reduces to

\begin{equation}
I_{\rm IB} = e \frac{R^S f^-_S R^T f^-_T}{R^S f^-_S  + R^T f^-_T}.
\label{eq:I_IB}
\end{equation}

Equation \eqref{eq:I_IB} tells us, even more clearly if cast into the form $I_{\rm IB}^{-1} = (eR^Sf_S^-)^{-1} + (eR^Tf_T^-)^{-1}$, that the current is the result of two competing processes happening \emph{in series}: the thermal unblocking of the decoupled state $|{\rm N}+1 \, E_1 {\mathrm d} \, \tau\rangle \to |{\rm N}\, E_0 \, 0\rangle$ towards the substrate and the tip tunnelling event $|{\rm N}+1 \, E_1 \,  {\rm c} \, \tau\rangle \to |{\rm N} \, E_0 \, 0\rangle$. Notice that in the system dynamics the two tunnelling events are not independent: one cannot happen if the other did not happen before.     
In the interference blocking regime $f^-_S \ll f^-_T$, but, in an STM set up, it typically also holds $R^T \ll R^S$. To fix the ideas let us first fix the tip position (thus, the ratio $R^T/R^S$) and lower the bias, deep in the interference blockade, such to fulfill the condition $R^S f^-_S \ll R^T f^-_T$. The current is thus proportional to $R^S$ and independent of the tip position. This fact explains the flattening of the constant height currnet map in Fig.~\ref{fig:Current_maps} and the wide plateau of the current versus tip molecule distance in Fig.~\ref{fig:Current_z}. Nevertheless, as the tip moves far of the molecule, the tip rate drops and, as the condition $R^T f^-_T \ll R^S f^-_S$ is fulfilled, the position dependence of the current is recovered ($I_{\rm IB} \propto R^T$). The cross over between the two regimes is estimated by the relation:
\begin{equation}
R^T(\vec{R}_{\mathrm{tip}},\Delta E) =
R^S{\mathrm e}^{\beta(\Delta E -  \mu_S)}.
\end{equation}
with $\beta = (k_{\rm B}T)^{-1}$. For completeness we add that the interference blockade is the only regime in which the current looses its canonical dependence on the tip position: the current saturates in fact to $I = -4eR^T$ for large positive biases, it is $I = -2eR^T$ at the tip-molecule resonance and $I = \frac{4}{5}eR^T$ at the substrate-molecule resonance. The summary of these results and their extension to the work function and bias voltage plane is presented in the left panel of Fig. \ref{fig:IV_stability} where the letters T and S indicate regions where the current is proportional respectively to the tip or substrate rate.

The enhanced sensitivity of the apparent molecular height to the value of the working current of a constant current scan performed in the interference blocking regime can also be explained by analyzing Eqs. \eqref{eq:I_CB} and \eqref{eq:I_IB}. Let us consider a certain value for the working current $I_0$. Starting from Eqs. \eqref{eq:I_CB} and \eqref{eq:I_IB} we can extract the equations for the constant current isosurfaces:

\begin{equation}
 \begin{split}
  R^T &= \frac{I_0}{4ef_S^+} \equiv K_{\rm CB}(I_0,V_b),\\ 
  R^T &= \frac{I_0}{ef_T^-}\left(1-\frac{I_0}{eR^Sf_S^-}\right)^{-1} \equiv K_{\rm IB}(I_0,V_b),
 \end{split}
 \label{eq:isosurface}
\end{equation}
respectively for the Coulomb blockade and interference blockade regimes. If, for a given choice of the parameters $I_0$ and $V_b$, it holds $K_{\rm IB} = K_{\rm CB}$ the two associated contant current isosurfaces coincide. This is indeed, by construction, the case for the bias corresponding to the points $2$ and $4$ in Fig.~\ref{fig:IV_stability} if the working current is chosen exactly as the one in the IV characteristics. Nevertheless, for the same choice of the biases, very different sensitivity of the constant current isosurface to the value of the working current is shown in the interference blockade and in the Coulomb blockade cases (compare upper to lower panel in Fig.~\ref{fig:Height_cuts}). By analyzing the second equation in \eqref{eq:isosurface} we can see that $K_{\rm IB}$ diverges for $I_0$ in the vicinity of the interference current $eR^Sf_S^-$ while $K_{\rm CB}$ shows a completely regular behaviour. As $K_{\rm IB} \to \infty$ the corresponding isosurface shrinks rapidly as it can be seen in Fig.~\ref{fig:Height_cuts}. Moreover, the interference 
current also represents in the vicinity of the interference blockade regime, an upper limit for the working current accessible to a constant current STM scan. In fact, for $I_0 > eR^Sf_S^-$ the constant $K_{\rm IB}$ turns negative and the second equation in \eqref{eq:isosurface} can not be fulfilled for whatever position of the tip.        

Special consideration should be given to the robustness of the presented effect. Indeed we have presented so far the idealized situation in which the rotational symmetry of the CuPc  is assumed to be unperturbed with a consequently perfect degeneracy of the anion ground states. Nevertheless this perfect degeneracy is not a necessary condition for the occurrence of the many-body interference effect described in the manuscript. As we have already explicitly shown in a previous publication \cite{DarauBDG_09} the interference blocking scenario persists as far as the quasi-degeneracy is present {\it i.e.} the splitting of the interfering energy levels is smaller than the tunneling coupling. In fact, if the tunneling coupling is strong enough, the indetermination principle does not allow to distinguish between the two quasi-degenerate states in the tunneling event and interference takes place. Moreover, since the tip tunneling coupling is controlled, in an STM experiment, by the tip position, the interference 
between quasi-degenerate states could be controlled by the tip position. The result would be the tuning, with the tip-molecule distance, of the negative differential conductance at negative bias voltages associated to the interference blocking. Finally, for what concerns the effect of the substrate on the molecular symmetry we would like to mention that strong experimental sensitivity to molecular symmetry has been proven for derivatives of CuPc molecules  on thin insulating films (see Sonnleitner et al. [\onlinecite{SonnleitnerSPPR_11}]) suggesting that an almost complete decoupling of the molecular states is indeed a good approximation for these systems.           

Finally, the results presented so far for the CuPc apply in general to the class of planar molecules belonging to the $C_{\mathrm{nv}}$ symmetry group, i.e. invariant under the set of rotations of angles $z 2\pi/n, \, z = 0,\ldots,n-1$ around a principal rotation axis perpendicular to the molecular plane and to a set of $n$ vertical planes (see Fig.~\ref{fig:Set-up}). Their many body-states, like the single particle ones, can be classified using the projection $\ell$ of the angular momentum in the direction of the main rotational axis (conventionally the $z$ axis) that we introduced in Eq.~\eqref{eq:H_tun}. The generic many body eigenstates of $H_{\mathrm m}$ can thus be written in the form $|{\rm N}\,E\,\ell\,S_z\rangle$, where $N$ is the particle number, $E$ the energy, $S_z$ and $\ell$ respectively the projections of the total spin and of the angular momentum in the $z$ direction in units of $\hbar$.  The state $|{\rm N}\,E\,\ell\,S_z\rangle$ transforms under a rotation of an angle $\phi = z 2\pi/n$ around the main 
rotation
axis as:
\begin{equation}
R_{\phi} |{\rm N}\,E\,\ell\,S_z\rangle= {\mathrm e}^{{\mathrm i }\phi(\ell + S_z)}|{\rm N}\,E\,\ell\,S_z\rangle.
\end{equation}
where $R_{\phi}$ is the rotation operator. Consequently it is not difficult to prove that $\ell$ is an integer number and $-\frac{n}{2}<\ell \leq \frac{n}{2}$ for $C_{\mathrm{nv}}$ molecules with even $n$ and $-\frac{n-1}{2}\leq \ell \leq \frac{n-1}{2}$ when $n$ is odd. Since $C_{\mathrm
{nv}}$ admits at maximum bidimensional irreducible representations, we conclude that the states with opposite $\ell$, connected by the reflection operation through the $n$ vertical planes, have symmetry protected degeneracy and only states with $\ell = 0$ (for even or odd $n$) or $\ell = 0,\frac{n}{2}$ (for even $n$) are non degenerate.

\section{Conclusions}

By studying the transport characteristics of an STM single molecule junction on a thin insulating film we identify in this article a class of molecules that should present strong NDC and interference blocking features. Moreover we establish a criterion to identify the interference blocking scenario based on topographical fingerprints. In particular, for biases in the vicinity of the interference blocking regime, a flattening of the molecular image in constant height and an enhanced sensitivity of the apparent height to the working current in the constant current mode are expected. The robustness of the effect is ensured by the observation that quasi-degeneracy and not exact degeneracy of the interfering many-body states is the necessary condition for the persistence of the phenomenon.

\begin{acknowledgments}
We thank prof.~Jascha Repp for fruitful discussions. Moreover, we acknowledge financial support by the DFG within the research programs SPP 1243, GRK 1570 and SFB 689.
\end{acknowledgments}

\appendix

 \section{Tunneling rates and overlap integrals}

The derivation of the tunneling rates, up to a number of small differences, is following the example given in\cite{SobczykDG_12}. In their most general form, they are given by:
\begin{subequations}
\begin{align}
\Gamma_{\ell m,\,\ell' m'}^{\chi} = \frac{2\pi}{\hbar}
&{}\sum_k\left(t^{\chi}_{k \ell m}\right)^*t^{\chi}_{k \ell' m'}
\delta(\varepsilon_k^{\chi}-\Delta E)\\
=\varepsilon_{\ell m}\varepsilon_{\ell' m'}
&{}\sum_{\alpha\beta} \langle \ell m\sigma | \alpha\sigma\rangle M_{\alpha\beta}^\chi \langle\beta\sigma | \ell'm'\sigma \rangle,
\end{align}
\end{subequations}
where
\begin{equation}
 M_{\alpha\beta}^\chi(\Delta E) = \frac{2\pi}{\hbar} \sum_{k} \delta(\varepsilon_k^\chi-\Delta E)
 \langle \alpha\sigma | \chi \vec k\sigma \rangle\langle \chi \vec k\sigma | \beta\sigma \rangle.
\end{equation}
The coefficients $\langle \ell m\sigma | \alpha\sigma\rangle$ and the energies $\varepsilon_{\ell m}$ are obtained by diagonalizing the single particle Hamiltonian of the molecule,
which is set up by using the Slater-Koster tight-binding approximation\cite{SlaterK_54}.
The state $|\alpha\sigma\rangle$ denotes an atomic orbital located at site $\alpha$ with position vector $\vec R_\alpha=(x_\alpha,y_\alpha,d)^\intercal$.
The corresponding wavefunctions are approximated by contracted Gaussian orbitals $g_{2p}(\vec r)$ and $g_{3d}(\vec r)$ to simplify the calculation of the overlap integrals. The definition of the Gaussian orbitals, their contraction coefficients $d_i,\,e_i$ and their exponents $a_i,\,b_i$ can be found in\cite{HehreSP_69,PietroLHS_80}. The orbitals used in this paper then are given by:
%
% \begin{equation}
%  p_{z,\alpha}(\vec r) = n_{\alpha}\sum_j(\vec r-\vec R_\alpha)\cdot\hat e_z\, d_{j,\alpha} \,
% \mathrm{e}^{-a_{j,\alpha}|\vec r-\vec R_\alpha|^2},
% \end{equation}
\begin{equation}
 p_{z}(\vec r) = n_{2p}\,\vec r\cdot\hat e_z\,g_{2p}(\vec r),
\end{equation}
for a $p_z$ orbital. A $d_{xz}$ orbital then accordingly reads:
%
% \begin{subequations}
%  \begin{align}
% d_{xz}(\vec r) =&{} n_{xz}\sum_j \vec r\cdot\hat e_x\, \vec r\cdot\hat e_z\, d_{j,xz} \mathrm{e}^{-a_{j,xz} |\vec r|^2}, \\
% d_{yz}(\vec r) =&{} n_{yz}\sum_j \vec r\cdot\hat e_y\, \vec r\cdot\hat e_z\, d_{j,yz} \mathrm{e}^{-a_{j,yz} |\vec r|^2}.
%  \end{align}
% \end{subequations}
 \begin{equation}
d_{xz}(\vec r) = n_{3d}\, \vec r\cdot\hat e_x\, \vec r\cdot\hat e_z\, g_{3d}(\vec r),
 \end{equation}
The parameters $n_{2p}$ and $n_{3d}$ are ensuring normalization.
The electronic states of the tip and the substrate are given by $|(\chi=T) \vec k\sigma\rangle$ or $|(\chi=S) \vec k\sigma\rangle$, respectively. Their wavefunctions can be expressed in the following form:
% In contrast to the latter, in the present publication only the exponentially decaying part
% of the wavefunction between the molecule and the substrate or the tip is used to evaluate the overlap integrals
% $\langle \chi \vec k\sigma | \alpha\sigma \rangle$. The leads' wavefunctions read:
%
\begin{equation}
 \Psi^\chi(x,y,z) = \psi^\chi_\parallel(x,y)\psi^\chi_\bot(z),
\end{equation}
where $\psi^\chi_\parallel(x,y)$ is given by plane waves for $\chi=\mathrm{S}$, or by the wavefunction of the groundstate of a twodimensional harmonic oscillator for $\chi=\mathrm{T}$. The wavefunctions $\psi^\chi_\bot(z)$ are the exponentially decaying parts of the solutions of one-dimensional finite potential wells:
% In contrast to\cite{SobczykDG_12}, for $\psi^\chi_\bot(z)$ only the exponentially decaying parts towards the molecule are used:
% %
% 
\begin{equation}
\psi^{\mathrm{S}}_\bot(z) = n_\bot^{\mathrm{S}}\,\mathrm{e}^{-\kappa_{\mathrm{S}} z} \quad\textmd{and}\quad
\psi^{\mathrm{T}}_\bot(z) = n_\bot^{\mathrm{T}}\,\mathrm{e}^{\kappa_{\mathrm{T}} (z-z_{\mathrm{tip}}) },
\end{equation}
where $n_\bot^\chi$ accounts for normalization and $\kappa_\chi$ is given by:
\begin{equation}
 \kappa_\chi=\sqrt{ \frac{2m}{\hbar^2}(-\varepsilon_0^\chi-\varepsilon_z) }.
\end{equation}
For the sake of reproduction, the different
contributions to $M_{\alpha\beta}^\chi$ are listed in the following.

\subsection{Substrate-molecule tunneling rates}

For two $p_z$ orbitals located at sites $\alpha$ and $\beta$, $M_{\alpha\beta}^{\mathrm S}$ reads:
%
% \begin{widetext}
\begin{align}
 M_{\alpha\beta}^{\mathrm S} = &{} \frac{4\pi^4}{\hbar^3} n_{2p}^2\sqrt{\frac{m^3}{2}}
\sum_{ij}\frac{d_{i} d_{j}}{a_{i} a_{j}}\nonumber\\
&{} \times \int_0^{\varepsilon_F^{\mathrm{S}}+\phi_0^{\mathrm{S}}} \frac{\mathrm{d}\varepsilon_z}{\sqrt{\varepsilon_z}}
J_0(\tilde{k}_{\mathrm{S}} |\vec R_{\alpha\beta}| )
\mathrm{e}^{ -\frac{\tilde k^2_{\mathrm{S}}}{4} ( a_{i}^{-1}  + a_{j}^{-1} )    }\nonumber\\
&{}  \times F(a_{i},\kappa_{\mathrm{S}},-d) F(a_{j},\kappa_{\mathrm{S}},-d),
\end{align}
where $J_n(x)$ is the $n$-th order Bessel function, $\vec R_{\alpha\beta}=\vec R_\alpha - \vec R_\beta$ and
$\tilde k_\mathrm{S}=\sqrt{\frac{2m}{\hbar^2}(\Delta E-\varepsilon_0^\mathrm{S}-\varepsilon_z)}$.
The function $F(a,\kappa,x)$ results from the overlap of $\psi_\bot^\chi(z)$ with an atomic Gaussian orbital and is given by:
%
% \begin{align}
% F(a,\kappa,x) = \frac{\kappa}{4}\sqrt{\frac{\pi}{a^3}}e^{-ax^2} \Bigg[ \nonumber\\
% \frac{2}{\kappa}\sqrt{\frac{a}{\pi}} -
% \exp\left( \frac{(\kappa+2ax)^2}{4a} \right) \mathrm{erfc}\left( \frac{\kappa+2ax}{2\sqrt{a}} \right)  \Bigg].
% \end{align}
\begin{equation}
F(a,\kappa,x) = \frac{n_\bot^\chi \mathrm{e}^{-ax^2}}{2a}
-\frac{n_\bot^\chi \kappa  }{4}\sqrt{\frac{\pi}{a^3}}\mathrm{erfc}\left( \frac{\kappa+2ax}{2\sqrt{a}} \right)\,
\mathrm{e}^{\kappa x + \frac{\kappa^2}{4a}}.
\end{equation}
% \end{widetext}
%
Here, erfc$(x)$ is the complementary error function. Consequently, it follows for a $d_{xz}$ orbital located at $\vec R_\alpha$ and a $p_z$ orbital at $\vec R_\beta$:
%
% \begin{widetext}
\begin{align}
M_{\alpha,xz;\beta}^{\mathrm{S}} = &{} -\frac{2\pi^4}{\hbar^3}\sqrt{\frac{m^3}{2}}n_{3d} n_{2p}\cos\theta_{\alpha\beta}
\sum_{ij}\frac{e_{i} d_{j}}{b_{i}^2 a_{j}} \nonumber\\
&{} \times  \int_0^{\varepsilon_F^{\mathrm{S}}+\phi_0^{\mathrm{S}}} \mathrm{d}\varepsilon_z \frac{\tilde k_{\mathrm{S}}}{\sqrt{\varepsilon_z}}
J_1(\tilde{k}_{\mathrm{S}} |\vec R_{\alpha\beta}|)
\mathrm{e}^{ -\frac{\tilde k^2_{\mathrm{S}}}{4} ( b_{i}^{-1}  + a_{j}^{-1} )    }\nonumber\\
&{} \times F(b_{i},\kappa_{\mathrm{S}},-d) F(a_{j},\kappa_{\mathrm{S}},-d),
\end{align}
where $\theta_{\alpha\beta}$ is the polar angle of the planar component of the vector $\vec R_{\alpha\beta}$.
From $M_{\alpha,xz;\beta}^{\mathrm{S}}$ one can obtain $M_{\alpha,yz;\beta}^{\mathrm{S}}$ by exchanging
the corresponding parameters and by replacing $\cos\theta_{\alpha\beta}$ with $\sin\theta_{\alpha\beta}$.
The expression $M_{\alpha,xz;\beta,yz}^{\mathrm{S}}$ vanishes exactly due to symmetry reasons, and finally
$M_{\alpha,xz;\beta,xz}^{\mathrm{S}}$ is given by:
\begin{align}
 M_{\alpha,xz;\beta,xz}^{\mathrm{S}}= &{} -\frac{\pi^4}{2\hbar^3}\sqrt{\frac{m^3}{2}}n_{3d}^2
\int_0^{\varepsilon_F^{\mathrm{S}}+\phi_0^{\mathrm{S}}}
\mathrm{d}\varepsilon_z \frac{\tilde k^2_{\mathrm{S}}}{\sqrt{\varepsilon_z}} \nonumber\\
&{}\times\left[
\sum_{j}\frac{e_{j}}{b_{j}^2}F(b_{j},\kappa_{\mathrm{S}},-d)
\mathrm{e}^{ -\frac{\tilde k^2_{\mathrm{S}}}{4b_{j} } }
\right]^2  \nonumber\\
\equiv &{} M_{\alpha,yz;\beta,yz}^{\mathrm{S}} .
\end{align}
\subsection{Tip-molecule tunneling rates}

Due to the fact that the planar energy component of the tip wavefunction is fixed at $\varepsilon_\parallel=\hbar\omega$, there is only  one single integration in energy to evaluate in order to obtain the tip-molecule tunneling rates.
Because of this, they are much more straightforward to calculate than their substrate-molecule counterparts: 
\begin{align}
 M_{\alpha\beta}^T =&
\frac{2\pi}{\hbar^2}\int_0^{-\varepsilon_0^\mathrm{T}} 
\mathrm{d}\varepsilon_zD(\varepsilon_z)
\langle \alpha\sigma | T \vec k\sigma \rangle\langle T \vec k\sigma | \beta\sigma \rangle
\delta(\varepsilon_k^\mathrm{T}-\Delta E)\nonumber\\
=&\frac{2\pi}{\hbar^2}\sqrt{\frac{m}{2}}\frac{L_{tip}}{\Delta E-\varepsilon_0^T-\hbar\omega} 
\langle \alpha\sigma | T \vec k\sigma \rangle\langle T \vec k\sigma | \beta\sigma \rangle.
\end{align} 
The parameter $L_{tip}$ stems from the one-dimensional density of states of the tip and it is cancelled later on by the normalization of the tip wavefunction.
Another effect of the single integration in energy is that all possible combinations of $p_z$, $d_{xz}$ and $d_{yz}$ orbitals are surviving. In order not to go beyond the constraints of this paper we only list the overlap integrals needed to construct the matrices $M^{\mathrm{T}}_{\alpha\beta}$.
After introducing the following parameters and abbreviations, $\nu^2=\frac{m\omega}{2\hbar}$, $\Delta y_\alpha=y_{\mathrm{tip}}-y_\alpha$, $\Delta x_\alpha=x_{\mathrm{tip}}-x_\alpha$ and finally $\kappa_\mathrm{T}=\sqrt{\frac{2m}{\hbar^2}(\hbar\omega-\Delta E)}$,
we are able to give the overlap integrals between the different orbitals located at $\vec R_\alpha$ and the tip wavefunction:
\begin{widetext}
\begin{align}
\langle \alpha\sigma | \mathrm{T} \vec k\sigma \rangle &= 
-n_{2p}\sqrt{2\pi}\nu\sum_j
\frac{ d_j}{a_j+\nu^2}
\exp\left(-\frac{\nu^2a_j}{\nu^2+a_j}(\Delta x_\alpha^2 + \Delta y_\alpha^2)\right)
 F(a_j,\kappa_{\mathrm{T}},d-z_{\mathrm{tip}})\\
\langle \alpha_{xz}\sigma | \mathrm{T} \vec k\sigma \rangle &= 
n_{3d}\sqrt{2\pi}\nu^3\sum_j
\frac{ e_j\,\Delta x_\alpha}{(b_j+\nu^2)^2} 
\exp\left(-\frac{\nu^2a_j}{\nu^2+a_j}(\Delta x_\alpha^2 + \Delta y_\alpha^2)\right)
 F(a_j,\kappa_{\mathrm{T}},d-z_{\mathrm{tip}})\\
\langle \alpha_{yz}\sigma | \mathrm{T} \vec k\sigma \rangle &= 
n_{3d}\sqrt{2\pi}\nu^3\sum_j
\frac{ e_j\,\Delta y_\alpha}{(b_j+\nu^2)^2} 
\exp\left(-\frac{\nu^2a_j}{\nu^2+a_j}(\Delta x_\alpha^2 + \Delta y_\alpha^2)\right)
 F(a_j,\kappa_{\mathrm{T}},d-z_{\mathrm{tip}}).
\end{align}
\end{widetext}

\end{document}